\documentclass[a4paper]{jpconf}

\RequirePackage{lineno,units,upgreek,wrapfig} 
\RequirePackage{graphicx,amsmath,amssymb,array,tabularx,url,xspace,subfig}

\newcommand*\patchAmsMathEnvironmentForLineno[1]{%
\expandafter\let\csname old#1\expandafter\endcsname\csname #1\endcsname
\expandafter\let\csname oldend#1\expandafter\endcsname\csname end#1\endcsname
\renewenvironment{#1}%
{\linenomath\csname old#1\endcsname}%
{\csname oldend#1\endcsname\endlinenomath}}%
\newcommand*\patchBothAmsMathEnvironmentsForLineno[1]{%
\patchAmsMathEnvironmentForLineno{#1}%
\patchAmsMathEnvironmentForLineno{#1*}}%
\AtBeginDocument{%
\patchBothAmsMathEnvironmentsForLineno{equation}%
\patchBothAmsMathEnvironmentsForLineno{align}%
\patchBothAmsMathEnvironmentsForLineno{flalign}%
\patchBothAmsMathEnvironmentsForLineno{alignat}%
\patchBothAmsMathEnvironmentsForLineno{gather}%
\patchBothAmsMathEnvironmentsForLineno{multline}%
}

\newcommand{\GeVoverc}{\ensuremath{\mathrm{GeV}\!/\mathit{c}}}

\newcommand{\pizero}{\pi^{0}}
\newcommand{\kzeroshort}{K^{0}_{\mathrm{s}}}

\newcommand{\vtwo}{v_{2}}

\newcommand{\directgamma}{\gamma,dir}
\newcommand{\decaygamma}{\gamma,bg}
\newcommand{\inclgamma}{\gamma,inc}
\newcommand{\vtwodirect}{v_{2}^{\mathrm{\directgamma}}}
\newcommand{\vtwodecay}{v_{2}^{\mathrm{\decaygamma}}}
\newcommand{\vtwoincl}{v_{2}^{\mathrm{\inclgamma}}}

\newcommand{\vtwopi}{v_{2}^{\mathrm{\pi^{\pm}}}}

\newcommand{\Nincl}{N^{\mathrm{\inclgamma}}}
\newcommand{\Ndecay}{N^{\mathrm{\decaygamma}}}
\newcommand{\Npizero}{N^{\mathrm{\pizero}}}

\renewcommand{\pt}{p_{\mathrm{T}}}
\newcommand{\mt}{m_{\mathrm{T}}}
\newcommand{\ket}{KE_{\mathrm{T}}}

\newcommand{\cmspbpblhc}{\sqrt{s_{\mathrm{NN}}}=\unit[2.76]{TeV}}
\newcommand{\cmsaarhic}{\sqrt{s_{\mathrm{NN}}}=\unit[0.2]{TeV}}

\makeatletter
\DeclareRobustCommand*{\bfseries}{%
  \not@math@alphabet\bfseries\mathbf
  \fontseries\bfdefault\selectfont
  \boldmath
}
\makeatother

\begin{document}

\title{Measurement of Direct-Photon Elliptic Flow in Pb-Pb Collisions at $\cmspbpblhc$}
\author{Daniel Lohner (for the ALICE Collaboration)}
\address{Physikalisches Institut, Ruprecht-Karls-Universit\"{a}t Heidelberg, Heidelberg, Germany}
\ead{Daniel.Lohner@cern.ch}

\begin{abstract}
We present the first measurement of the direct-photon elliptic flow $\vtwodirect$ in Pb-Pb collisions at
 $\cmspbpblhc$ with data taken by the ALICE experiment at the LHC. The measurement provides evidence for a non-zero $\vtwodirect$ for $\unit[1 < \pt < 3]{\GeVoverc}$ with a magnitude similar to the observed charged pion $\vtwopi$. In order to explain the large inverse slope parameter $T_{\mathrm{eff}}$ of the low $\pt$ direct-photon spectrum observed at LHC and RHIC, recent hydrodynamical descriptions of the direct-photon production include a substantial portion of thermal photons from the hot plasma phase. As a consequence of the early production time, $\vtwodirect$ is expected to be small compared to hadrons. A large $\vtwodirect$ might lend support for a significant direct-photon emission from late stages of the system evolution where hadron flow has developed. 
\end{abstract}

\section{Introduction}
\begin{wrapfigure}{r}{0.5\textwidth}
  \begin{center}
    \includegraphics[width=0.48\textwidth]{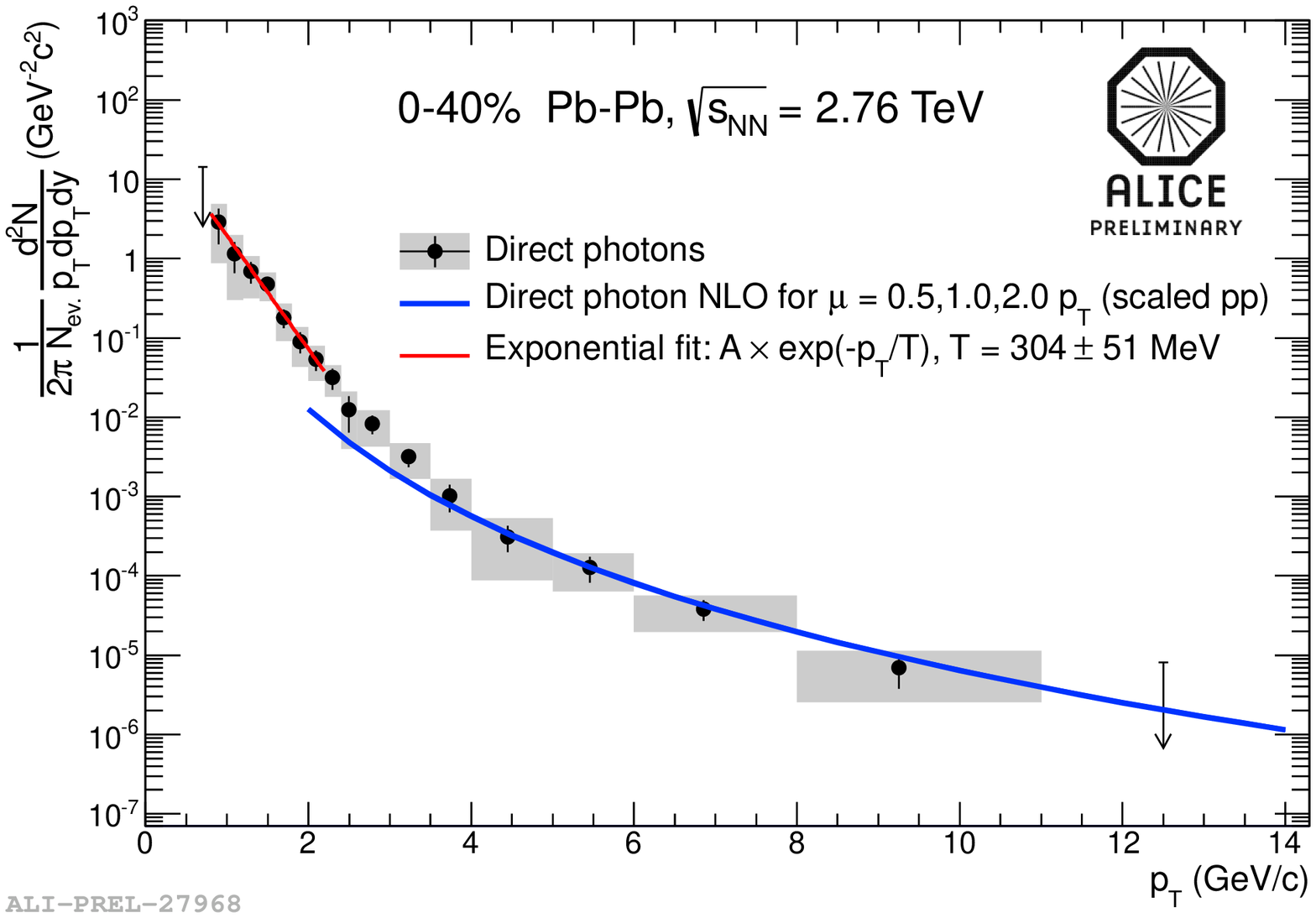}
  \end{center}
  \caption{Direct-photon invariant yield in $\unit[0\text{--}40]{\%}$ Pb-Pb collisions with NLO pQCD predictions and exponential fit \cite{Wilde:2012wc}.}
  \label{fig:directphotonspec}
\end{wrapfigure}

A unique tool for the study of the collision evolution in nucleus-nucleus collisions is the measurement of photons. Besides photons from hadron decays also direct photons are emitted at every stage of the system evolution. Since photons interact only weakly with the strongly coupled medium they carry undistorted information of the system at their production time \cite{PhysRevLett.96.202302}. \\
Theoretical descriptions of direct-photon production in nucleus-nucleus collisions include prompt photons, fragmentation photons, thermal photons and photons from parton-medium interactions \cite{Liu:2007tw,Gale:2009gc}.
Prompt photons originate from primary, hard interactions of partons including quark-antiquark-annihilation ($q+\bar{q} \rightarrow g+\gamma$) and quark-gluon compton scattering ($q+g \rightarrow q+\gamma$). The production of prompt photons can be described by next-to-leading-order (NLO) perturbative QCD. 
Fragmentation photons are produced in the fragmentation of hard scattered quarks or gluons (e.g. $q+q\rightarrow q+q+\gamma$).
Thermal photons are emitted by the hot thermalized medium through scattering of particles (e.g. $q+\bar{q} \rightarrow g+\gamma$) during the QGP phase and hadronic interactions (e.g. $\pi^{+}+\pi^{-}\rightarrow \rho_{0}+\gamma$) in the hot hadron gas phase.\\
The direct-photon spectrum can be calculated from the measured inclusive photon spectra by subtraction of all contributions from hadron decays. Measurements at RHIC \cite{Adare:2008ab} and LHC \cite{Wilde:2012wc} yield a significant contribution of direct photons to the inclusive photon spectra. Figure \ref{fig:directphotonspec} shows the direct-photon spectrum in the $\unit[0\text{--}40]{\%}$ most central Pb-Pb collisions at $\cmspbpblhc$. At transverse momenta above $\unit[4]{\GeVoverc}$ the signal is described by binary scaled NLO (pQCD) calculations for pp  \cite{Wilde:2012wc}. 
At low $\pt$ the spectrum shows a significant excess above the pQCD prediction which can be described by an exponential function with an inverse slope parameter of  $T^{\mathrm{LHC}}_{\mathrm{eff}}=\unit[(304 \pm 51^{stat+sys})]{MeV}$ \cite{Wilde:2012wc}.

The PHENIX collaboration reports an inverse slope parameter of  $T^{\mathrm{RHIC}}_{\mathrm{eff}}=\unit[(220 \pm 19^{stat} \pm 19^{sys})]{MeV}$ in $\unit[0\text{--}20]{\%}$ Au-Au collisions at $\cmsaarhic$ \cite{Adare:2008ab}. Assuming a thermal origin of low $p_T$ direct photons the inverse slope parameter can be interpreted as an effective temperature of the source integrated over the whole system evolution. A high effective temperature compared to the critical temperature for deconfinement ($\unit[150\text{--}170]{MeV}$ \cite{Aoki:2009sc,Cheng:2006qk}) implies an early production time of direct photons during the system evolution.\\
A fingerprint of the thermalization of the system created in nucleus-nucleus collisions is the observation of azimuthally anisotropic particle emission. The azimuthal anisotropy is studied via a Fourier decomposition with harmonic coefficients $v_{n}$. In non-central collisions the initial asymmetry of the overlap zone leads to a dominant second harmonic $v_{2}$ known as elliptic flow. At small transverse momenta ($\pt \lesssim \unit[3]{\GeVoverc}$) the hadron $\vtwo$ can be understood in terms of the collective expansion of the medium. At higher momenta the anisotropy of hadrons is thought to be caused by the in-medium energy loss of partons.  \\ 
While measurements of hadrons quantify the system evolution at the kinetic freeze-out, direct photons carry the system information at their production time. Consequently, a measurement of $\vtwodirect$ allows to put additional constraints on their production time. Hydrodynamical descriptions of thermal photon production include a substantial portion of photons emitted in the hot plasma phase where flow has not yet been developed. Other sources of direct photons have either small or zero elliptic flow. Thus, a generic model expectation is that the $\vtwodirect$ is smaller than for charged particles \cite{Holopainen:2011pd,vanHees:2011vb}.

\section{Analysis Method}
The analysis is based on about $2 \times 10^{7}$ minimum bias Pb-Pb collisions at $\cmspbpblhc$ recorded in 2010 with the ALICE experiment at the LHC.
Photons were detected at mid rapidity ($\lvert\eta\rvert \leq 0.8$) via their conversion in the detector material using the tracking and particle identification capabilities
of the Inner Tracking System (ITS), the Time Projection Chamber (TPC) and the Time-Of-Flight (TOF) detector. Photon candidates were reconstructed using a Kalman filter based secondary vertex finding algorithm. A set of selection criteria on kinematic variables and particle identification allows to separate photons efficiently from other sources of secondary vertices (combinatorial background, $\Lambda$, $\kzeroshort$, etc)  \cite{Abelev:2012cn}. Centrality and event plane angle were determined using the VZERO detector ($-3.7 \leq \eta \leq -1.7$ and  $ 2.8 \leq \eta \leq 5.1$).\\
The inclusive photon $\vtwoincl$ was studied in small centrality classes in order to minimize systematic multiplicity effects ($\unit[0\text{--}5]{\%}, \unit[5\text{--}10]{\%}, \unit[10\text{--}20]{\%}, \unit[20\text{--}30]{\%}, \unit[30\text{--}40]{\%}$). The direct-photon spectrum was measured in the $\unit[0\text{--}40]{\%}$ centrality class \cite{Wilde:2012wc}. In order to minimize systematic uncertainties the ratio $\frac{\Nincl}{\Ndecay}$ was calculated via the double ratio $R$:
\begin{align}
 R= \frac{\left(\frac{d\Nincl/d\pt}{d\Npizero/d\pt}\right)}{\left(\frac{d\Ndecay/d\pt}{d\Npizero/d\pt}\right)_{MC}} = \frac{\Nincl}{\Ndecay}
\end{align}
The inclusive photon spectrum was obtained from Pb-Pb data. The decay photon spectrum was calculated in a cocktail simulation based on the measured neutral pion spectrum. The contribution of $\eta$'s and other mesons was estimated by transverse mass ($\mt$) scaling.
In order to estimate the decay photon elliptic flow $\vtwodecay$, the initial hadron azimuthal distributions were parametrized with the measured charged pion $\vtwopi$ \cite{ALICEFlow} scaled in transverse kinetic energy ($\ket = \mt -m_{0}$). Finally, the direct-photon $\vtwodirect$ was calculated from the double ratio $R$, inclusive photon $\vtwoincl$ and decay photon $\vtwodecay$:
\begin{align}
\label{eq:directv2}
\vtwodirect=\frac{R\vtwoincl-\vtwodecay}{R-1}
\end{align}

\section{Results}
\begin{figure}[h!!!]
   \centering
\begin{minipage}[l]{0.45\linewidth}
\centering
\includegraphics[width=75mm]{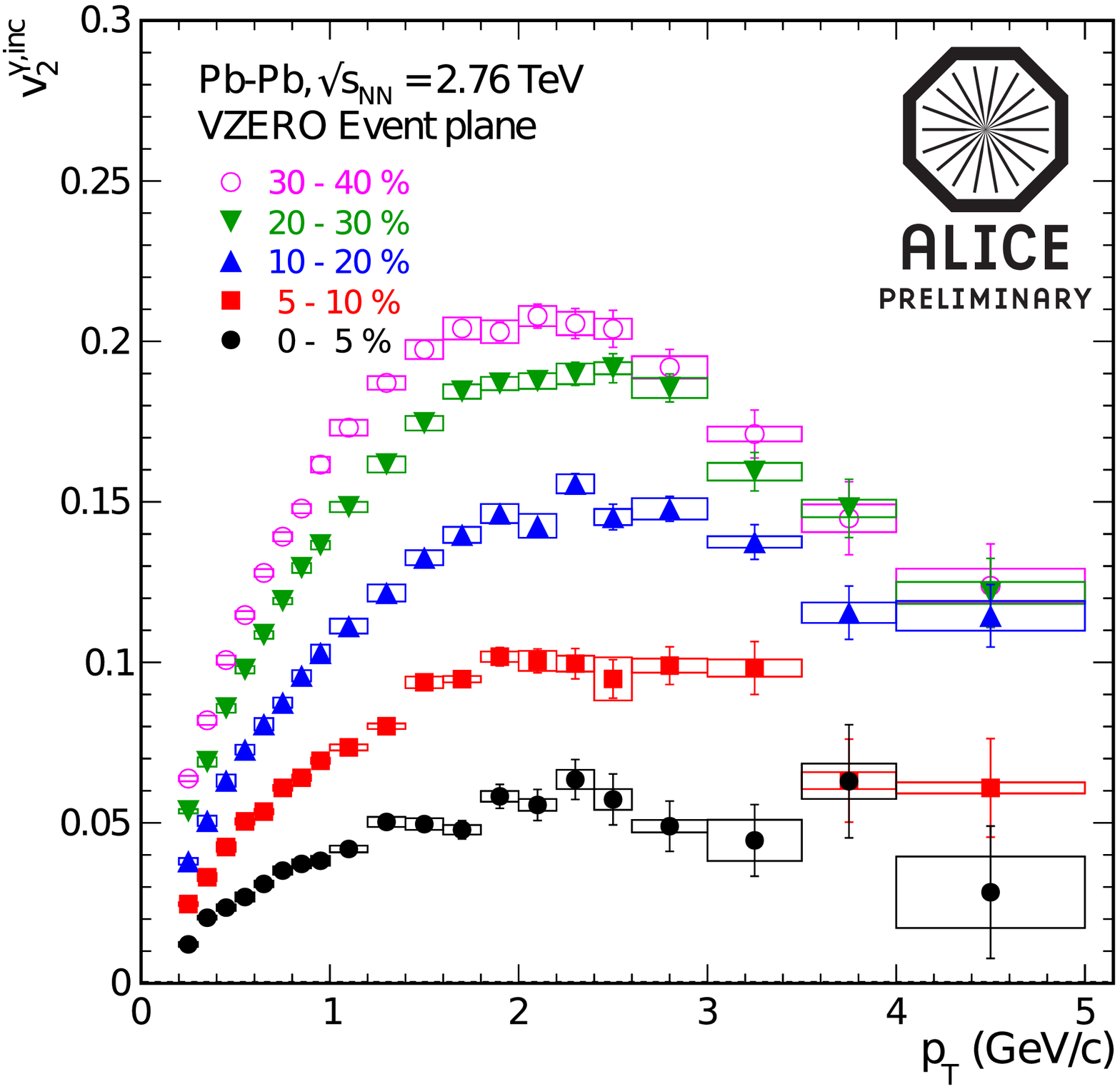}
\caption{Inclusive photon $\vtwoincl$ for several centrality classes in Pb-Pb collisions.}
\label{fig:inclusivev2}
\end{minipage}
\hspace{0.5cm}
\begin{minipage}[c]{0.45\linewidth}
\centering
\includegraphics[width=75mm]{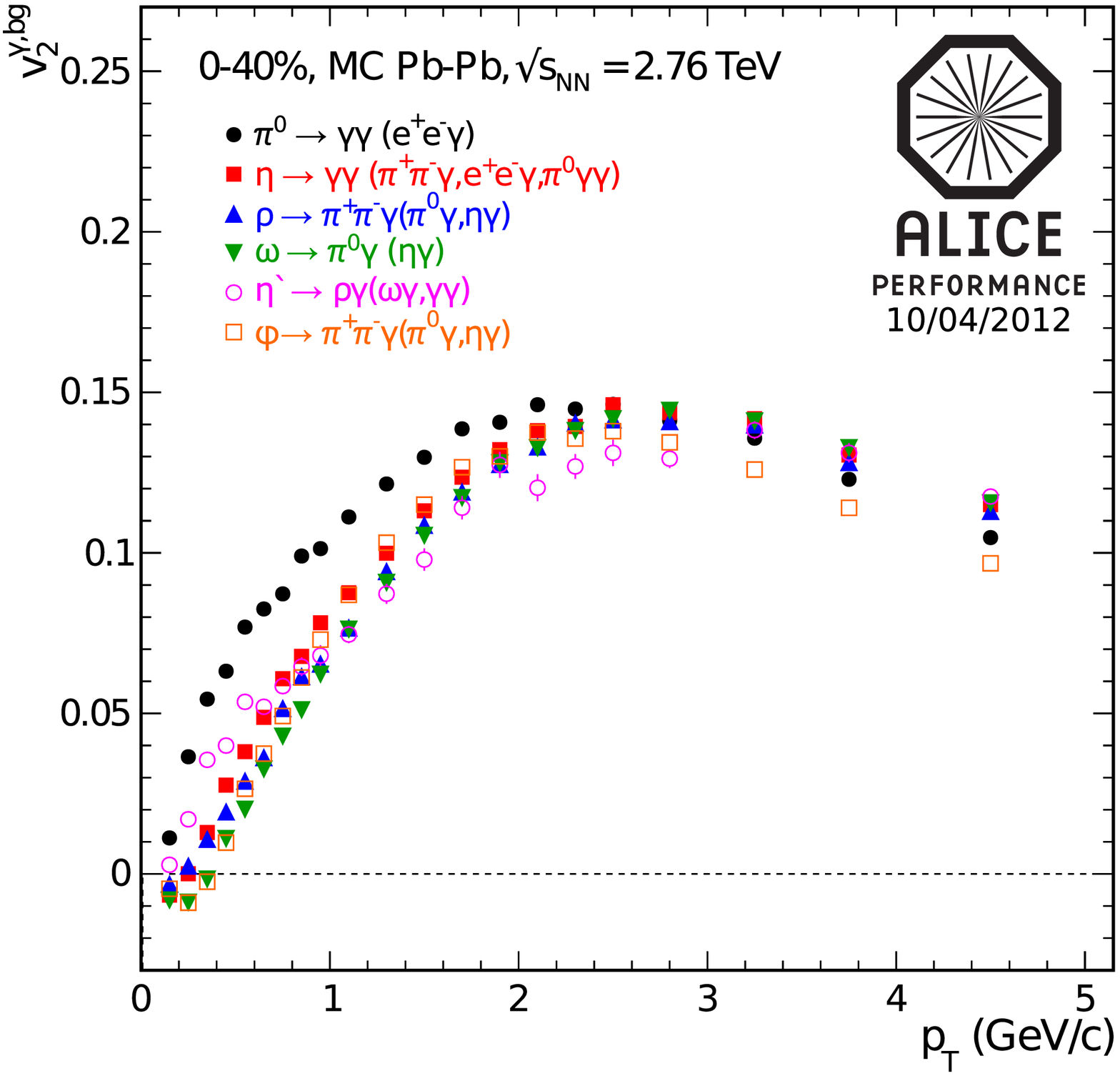}
\caption{Simulation of decay photon $\vtwodecay$ from different hadron decays.}
\label{fig:cocktailphotonv2}
\end{minipage}
 \end{figure}

Figure \ref{fig:inclusivev2} shows the inclusive photon $\vtwoincl$ for several centrality classes in Pb-Pb collisions. The magnitude of $\vtwoincl$ is comparable to hadronic $v_{2}$ \cite{ALICEFlow} and decreases with increasing centrality. 
Figure \ref{fig:cocktailphotonv2} shows the $v_{2}$ components for decay photons from various hadrons. The magnitude of decay photons is also comparable to hadrons but the values are shifted in transverse momentum $\pt$ depending on the mass of the mother particles. The decay kinematics can even result into negative values for $v_{2}$ at small transverse momenta. About $80\%$  of the decay photons come from neutral pion decays ($\pi^{0}\rightarrow \gamma \gamma$) and about $\unit[18]{\%}$ from  the $\eta$ meson ($\eta \rightarrow \gamma \gamma$).

\begin{figure}[h!]
   \centering
\begin{minipage}[l]{0.45\linewidth}
\centering
\includegraphics[width=75mm]{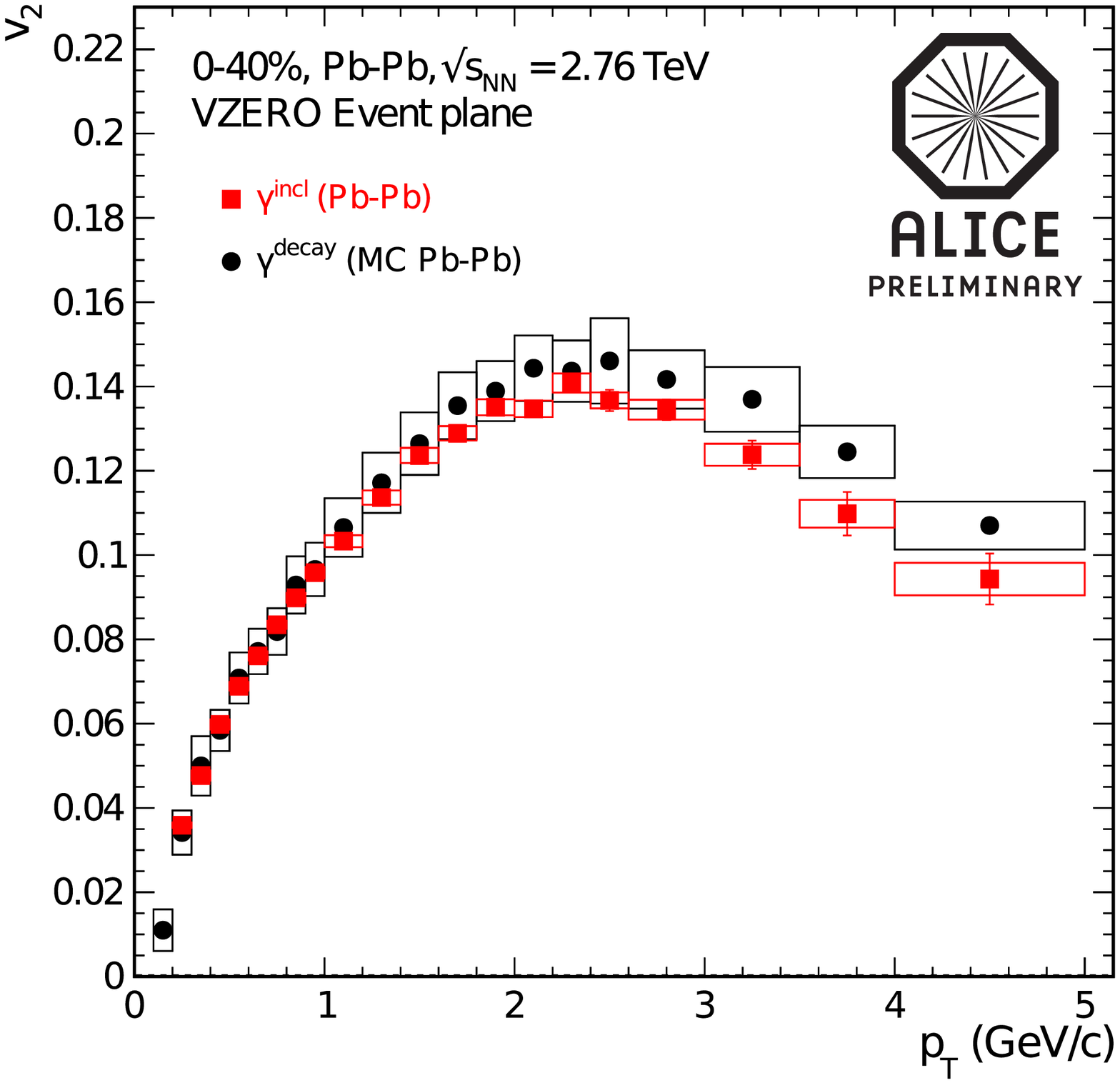}
\caption{Inclusive photon $\vtwoincl$ and decay photon $\vtwodecay$ in $\unit[0\text{--}40]{\%}$ Pb-Pb collisions.}
\label{fig:comparison}
\end{minipage}
\hspace{0.5cm}
\begin{minipage}[c]{0.45\linewidth}
\centering
\includegraphics[width=75mm]{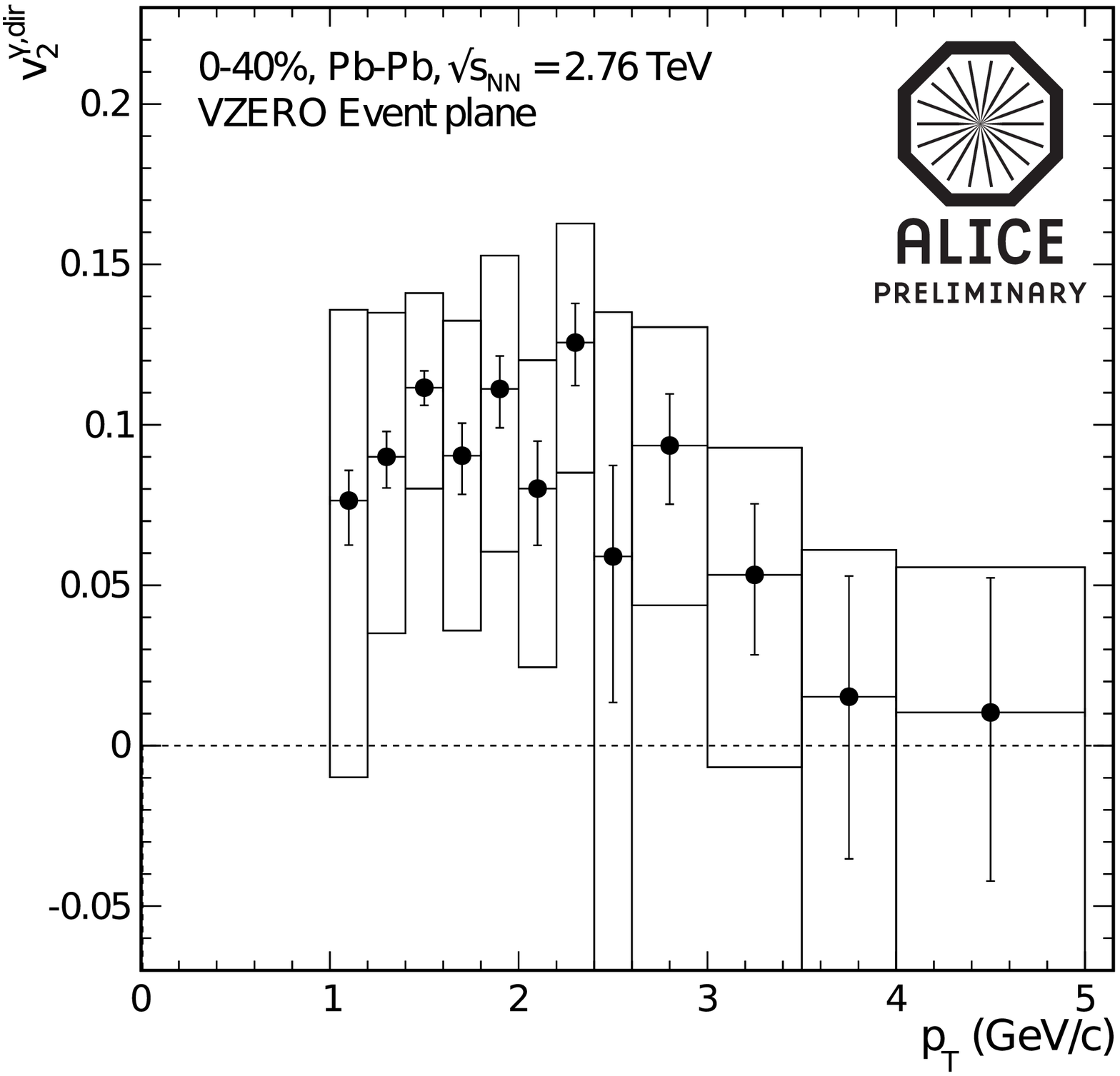}
\caption{Direct-photon $\vtwodirect$ in $\unit[0\text{--}40]{\%}$ Pb-Pb collisions.}
\label{fig:directphotonv2}
\end{minipage}
 \end{figure}

 The accumulated $\vtwodecay$ is shown in Fig.~\ref{fig:comparison} in comparison to the inclusive photon $\vtwoincl$ in $\unit[0\text{--}40]{\%}$. Below a transverse momentum $\pt$ of \unit[3]{\GeVoverc} decay and inclusive photons are consistent within uncertainties which allows for two interpretations: either direct and decay photons have a similar elliptic flow ($\vtwodirect \approx \vtwodecay$) or there are no direct photons ($R=1$). Above \unit[3]{\GeVoverc} the inclusive photon $\vtwoincl$ is significantly smaller than $\vtwodecay$ which can only be explained by a direct-photon contribution with smaller $\vtwodirect$ compared to $\vtwodecay$.
Systematic uncertainties of the decay photon $\vtwodecay$ include very conservative assumptions for a breaking of $\mt$ and $\ket$ scaling, a possible deviation between neutral and charged pion $v_{2}$ and the unknown centrality dependence of direct-photon production. Consequently, the estimate of the $\vtwodecay$ is the dominant contribution of systematic uncertainty of the $\vtwodirect$ measurement.\\
Figure \ref{fig:directphotonv2} shows the first measurement of $\vtwodirect$ in  $\unit[0\text{--}40]{\%}$ Pb-Pb collisions at $\cmspbpblhc$.

\section{Summary and Conclusions}
The results provide evidence for a non-zero $\vtwodirect$  for $\unit[1 < \pt < 3]{\GeVoverc}$ with a magnitude similar to the observed charged pion $\vtwopi$ \cite{ALICEFlow}. Similar results were reported by the PHENIX collaboration \cite{Adare:2011zr}. Recent hydrodynamical calculations \cite{Holopainen:2011pd,vanHees:2011vb} include a substantial portion of thermal photons from the hot plasma phase and also a sizable fraction from other sources in order to describe the observed direct-photon spectra. However, the emission from early stages of the system evolution yields a small $\vtwodirect$ compared to hadrons. Thus, the observed large $\vtwodirect$ might lend support for a significant emission from late stages of the system evolution where the hadron flow has built up.

\section*{References}

\bibliographystyle{iopart-num}
\bibliography{../bibtex/literature}


\end{document}